\input{aipcheck}

\documentclass[
    ,final            
  ]
  {aipproc}

\layoutstyle{8x11single}

\usepackage{graphics}
\usepackage{graphicx}
\usepackage{epsfig}
\usepackage{epstopdf}

\begin{document}

\title{Open Heavy Flavor Production in p-p and Pb-Pb collisions as seen by ALICE at the LHC}

%



\author{Zaida Conesa del Valle, for the ALICE Collaboration}{
  address={Organisation Europ\'eenne pour la Recherche Nucl\'eaire (CERN), Geneva, Switzerland}
}

\begin{abstract}
Charm and beauty production are probed with the ALICE experiment at the LHC by studying the single lepton transverse momentum distribution (electrons at mid-rapidity, muons at large-rapidities) and D mesons reconstructed in their hadronic decays. 
The differential production cross sections in proton proton interactions show a good agreement with perturbative QCD calculations at both $\sqrt{s}= 2.76$ and $7$~TeV. 
The measurements in lead lead reactions at $\sqrt{s_{\rm NN}}= 2.76$~TeV evidence a reduction (or suppression) of the production rate at intermediate and high $p_t$ in the most central collisions with respect to the rate in proton proton interactions. 
\end{abstract}

\maketitle


\section{Introduction}

The ALICE experiment has the ability to measure, in addition to other observables, open heavy flavor production in different colliding systems at the LHC. Charm and beauty production in proton proton collisions at the LHC are an important tool to test pQCD calculations in a new energy domain. Their spectra in heavy ion interactions are influenced by the formation of hot and dense QCD matter. Due to their relative large mass and the difference between quark and gluon color charge, the medium effects in heavy flavor hadrons are expected to differ from those of light hadrons. RHIC results show that heavy quarks effectively lose energy in the medium through (elastic and/or inelastic) interactions, but a possible quark-mass hierarchy has not yet been elucidated. LHC measurements at higher energies, with larger cross sections and better capabilities to separate charm and beauty production, will help to answer these questions. 

The status of open heavy flavor measurements in proton proton (pp) collisions at $\sqrt{s}=7$ and 2.76 TeV and lead lead (Pb-Pb) collisions at  $\sqrt{s_{NN}}=2.76$ TeV with the ALICE experiment is described. The results of the heavy flavor lepton analysis (electrons at mid-rapidity, muons at large-rapidities) and of D mesons reconstructed in their hadronic decays are discussed.

\section{Measuring heavy flavor production in ALICE}

The ALICE experiment consists of a central barrel ($|\eta|<0.9$), a muon spectrometer ($-2.5>\eta>-4$) and a set of detectors for trigger and event characterization purposes. It is characterized by its ability to track particles from very low ($\sim100$~MeV/c) up to fairly high ($\sim100$~GeV/c) transverse momenta. 
The central barrel, equipped with a solenoid magnet of up to $0.5$~T, is in charge of tracking and identifying charged particles and photons. 
From the vertex region to the outer part it is composed of a vertex detector, the Inner Tracking System (ITS), a large Time Projection Chamber (TPC), a Transition Radiation Detector (TRD) for electron identification, and a Time Of Flight detector (TOF) to identify protons, kaons and pions, all of them with full azimuthal acceptance. 
The muon spectrometer is equipped with a dipole magnet of up to $0.7$~T, and is responsible for muon tracking and reconstruction. 
The V0 detector comprising two scintillator hodoscopes located in the forward and backward regions, is involved in fast triggering and centrality determination. 
The minimum bias trigger is activated by the presence of at least a signal in either of two V0 scintillator hodoscopes, or in the two inner layers of the ITS. 
The experiment is completed by a set of global and central detectors that are not used in the measurements presented here.

Electrons were studied for $|\eta|<0.8$, reconstructed with the central tracking system (ITS, TPC), and identified thanks to the energy deposit in the TPC, the timing in the TOF, and the information in the TRD (TRD was only used in the pp sample). The heavy flavor electron spectrum was obtained subtracting a cocktail of the non-heavy-flavor sources from the inclusive electron spectrum and applying the acceptance and efficiency corrections. 
The cocktail was built based on the measured $\pi^0$, ${\rm J}/\psi$ and $\Upsilon$ spectrum (the last two only in pp), heavier meson spectra from the measured $\pi^0$ via $m_T$ scaling, and direct $\gamma$ from NLO pQCD calculations. 
In pp collisions, beauty decay electrons were identified based on the longer flight time of beauty hadrons, considering a constraint on the electron origin distance to the event primary vertex position (see details in~\cite{Dainese:2010ms}).

Muons were reconstructed with the muon tracking chambers ($-2.5>\eta>-4$) and identified by requiring the tracks to pass through the muon iron filter and reach the muon trigger chambers. The heavy flavor muon differential distribution was then built by subtracting statistically muons from hadronic decays (mainly from secondary decays of $K$ and $\pi$) from the inclusive muon spectra (details in~\cite{Dainese:2010ms}) and correcting for the acceptance and efficiency.

D mesons were reconstructed in their hadronic decays : ${\rm D}^0 (\bar{{\rm D}^{0}}) \rightarrow K \, \pi$, ${\rm D}^{\pm} \rightarrow K \, \pi \, \pi$, ${\rm D}^{*\pm} \rightarrow {\rm D}^0 (\bar{{\rm D}^{0}}) \, \pi$ by identifying their decay products ($K^{\pm}$, $\pi^{\pm}$) and applying topological constrains on the decay kinematics (candidate decay length, distance and angles of the decay products, \dots). 
The prompt charm distributions ($c \rightarrow {\rm D}$) were determined in $|y|<0.5$ after subtracting the feed-down contribution ($b \rightarrow {\rm D}$), based on FONLL pQCD calculations (details in~\cite{Dainese:2010ms}), and applying the acceptance and efficiency corrections.
%

\section{Results}

\begin{figure}[!htbp]
  \includegraphics[width=0.5\textwidth]{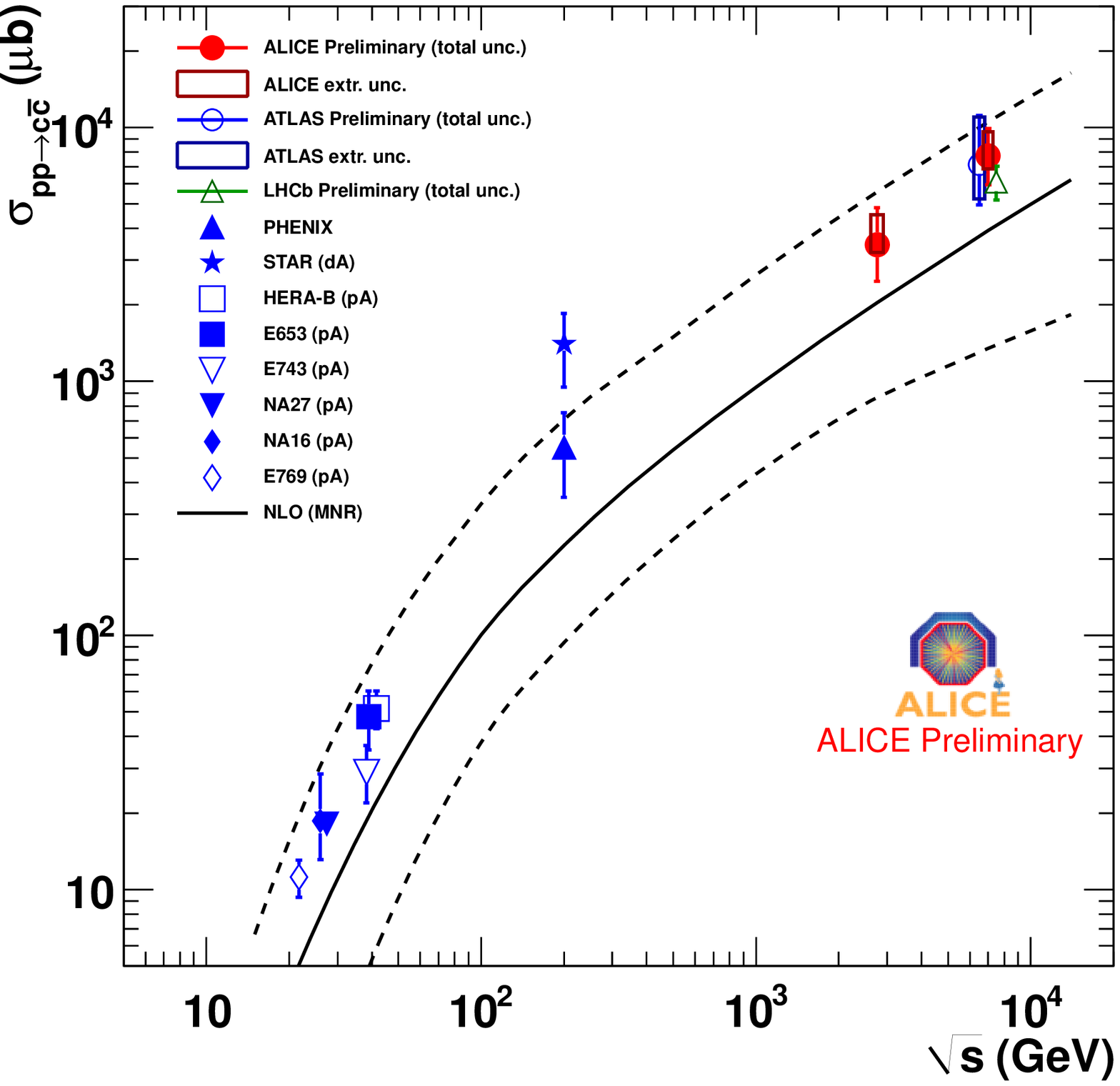}
  \includegraphics[width=0.425\textwidth]{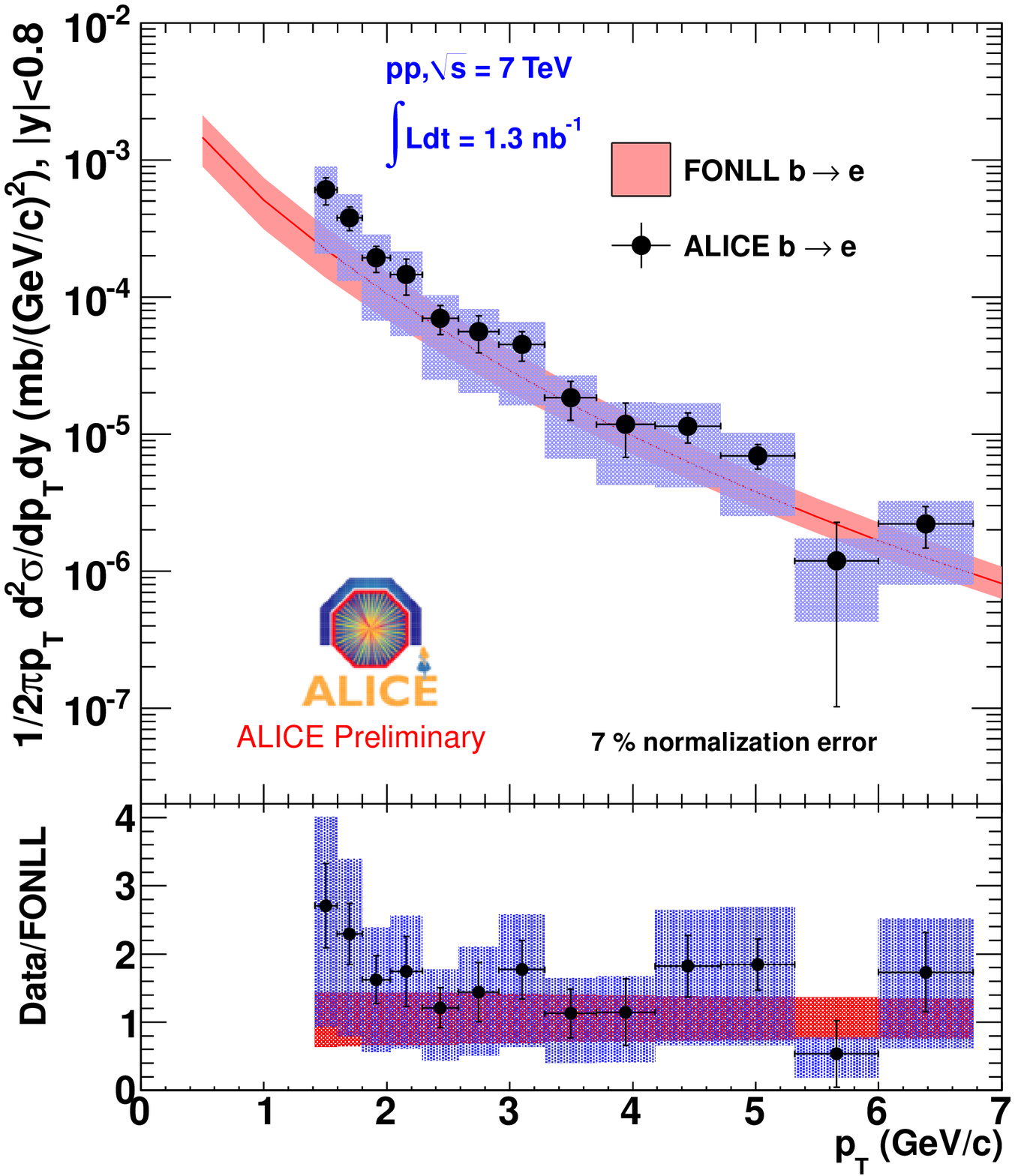}
  \caption{
  	\label{fig:ALICEHFpp}
  	Charm and beauty production cross section in proton proton collisions.
  	Left: Total charm cross section dependence with the system center of mass energy. 
	Right: Cross section of electrons from beauty hadron decays at 7~TeV. }
\end{figure}

The heavy flavor electron differential $p_t$ cross section and the heavy flavor muon $\eta$ and $p_t$ differential spectra were measured in pp collisions at $7$~TeV and shown to be in agreement with theoretical FONLL pQCD predictions in different pseudo-rapidity ranges~\cite{Dainese:2010ms}. 
Prompt D mesons were studied at both $7$~TeV and $2.76$~TeV in pp interactions. Their transverse momentum distributions were also well described by the existent pQCD calculations (FONLL~\cite{FONLL} and GM-VFNS~\cite{GMVFNS}) at both energies, the measurement and theoretical uncertainties not allowing to favor any of them.
The extrapolation of the D meson measurements (down to $p_t=2$~GeV/c) to the full kinematic phase space allowed to evaluate the total charm production cross section. Its dependence with the system center of mass energy, shown in Fig.~\ref{fig:ALICEHFpp}~left, illustrates the compatibility of the different experiments, and the seemingly good description of its energy evolution by the MNR pQCD calculations~\cite{MNR}. 
The beauty decay electron differential $p_t$ spectrum at $7$~TeV (see Fig.~\ref{fig:ALICEHFpp} right) also presents a good agreement with FONLL pQCD calculations. 

\begin{figure}[!htbp]
\centering
  \includegraphics[width=0.55\textwidth,height=0.45\textwidth]{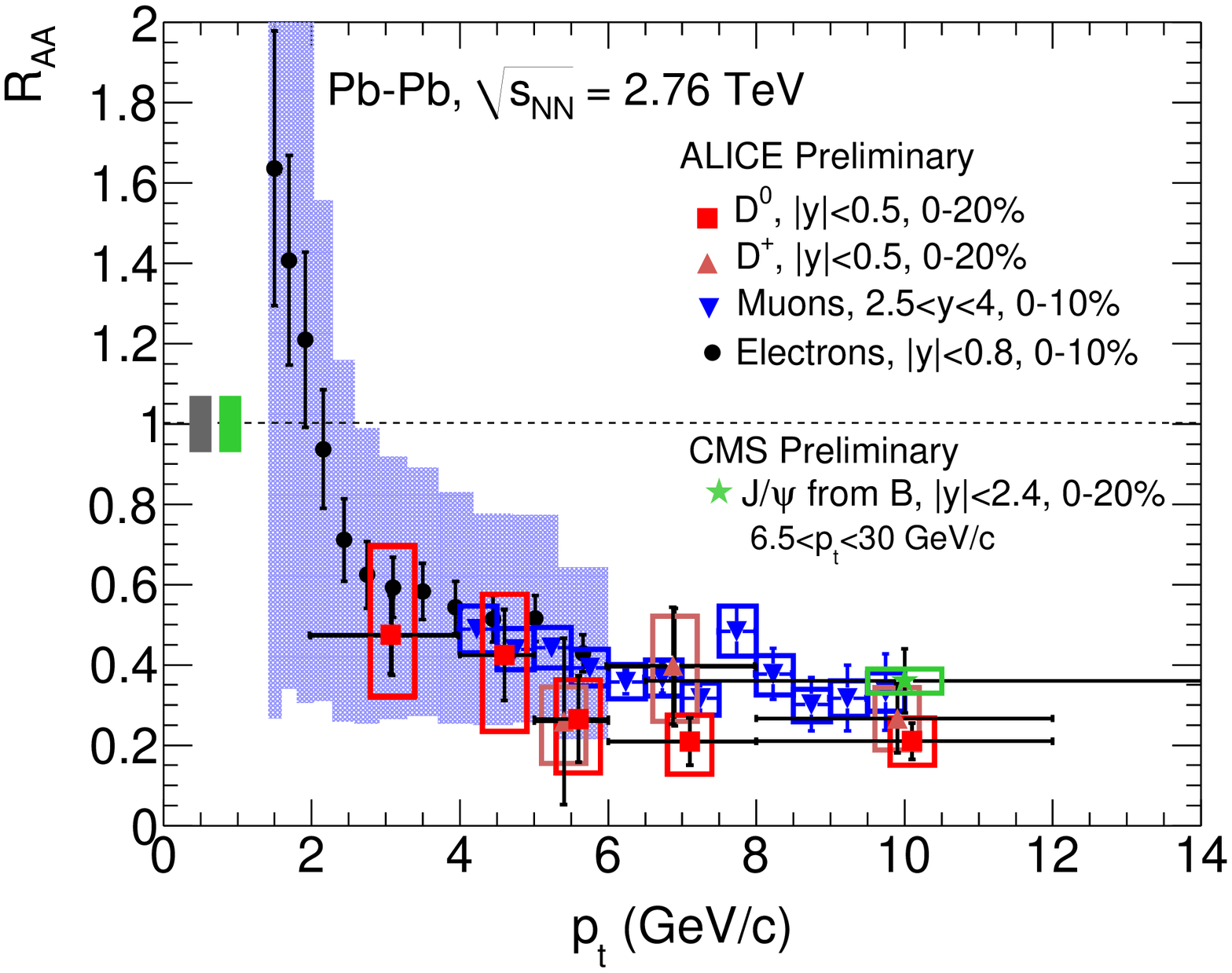}
  \includegraphics[width=0.5\textwidth,height=0.4\textwidth]{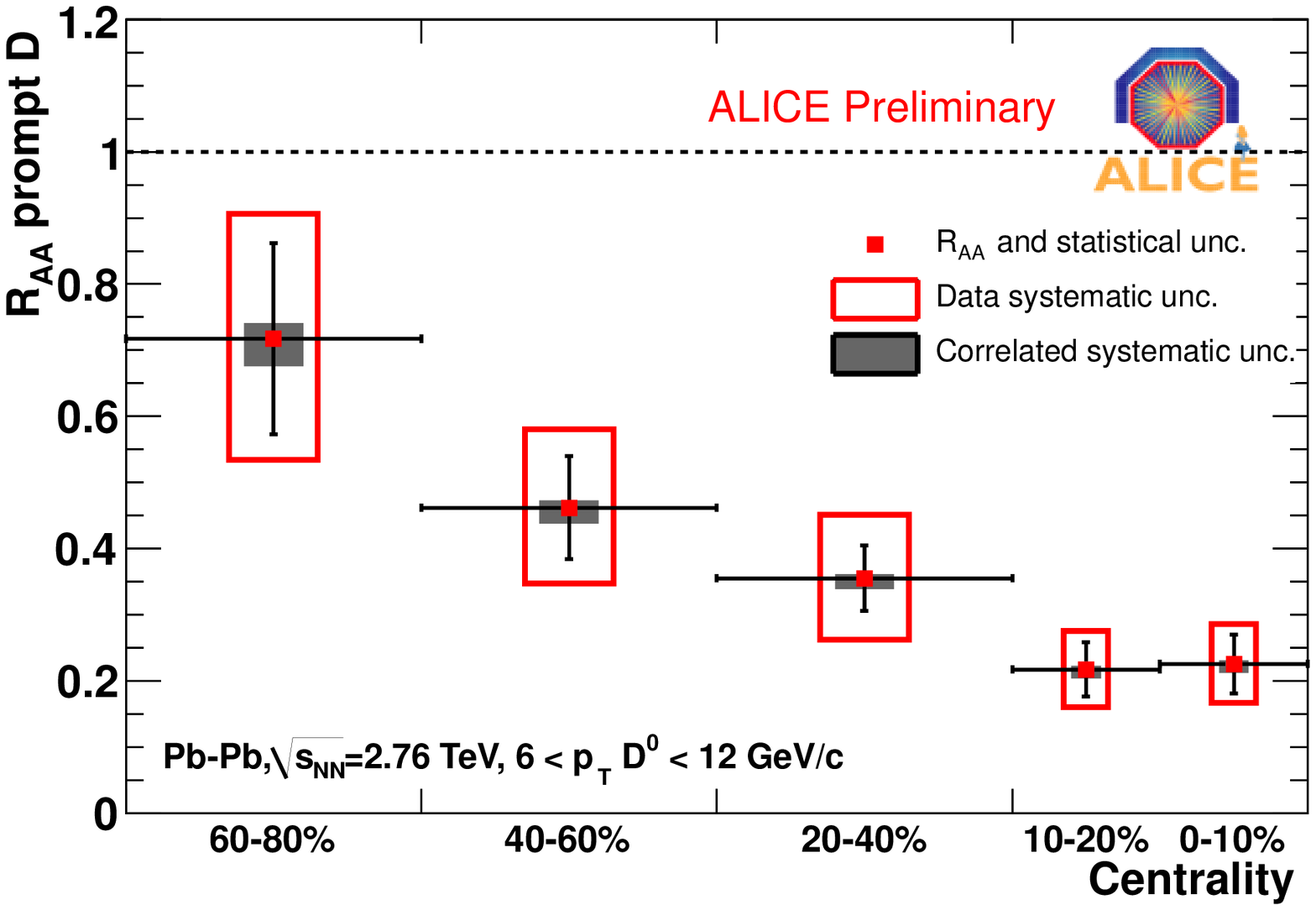}
  \caption{
  	\label{fig:ALICEHFPbPb}
  	Nuclear modification factor ($R_{\rm AA}$) of heavy flavor probes in Pb-Pb collisions at 2.76~TeV.
  	Left: comparison of the measurements in the most central collisions. 
	Right: $D^{0}$-meson $R_{\rm AA}$ centrality dependence for $6<p_t<12$~GeV/c at $|y|<0.5$.}
\end{figure}

Open heavy flavor production was also studied in PbPb reactions at $\sqrt{s_{NN}}=2.76$~TeV with the data sample collected in 2010~\cite{Dainese:2010ms}. The influence of the hot and dense QCD medium created in these heavy ion interactions was probed computing the nuclear modification factor, ${\rm R}_{\rm AA} = Y_{\rm AA} / ( \langle N_{\rm coll}\rangle \times Y_{\rm pp} )$, as the ratio of the invariant yields in the PbPb and pp collisions at the same energy normalized by the average number of nucleon nucleon collisions (see Fig.~\ref{fig:ALICEHFPbPb}). 
For all the herein mentioned measurements (single muons, single electrons, D mesons), the pp reference was obtained by extrapolating the $7$~TeV measurements presented in the previous section to $2.76$~TeV~\cite{Averbeck:2011ga}. The scaling factor was calculated from pQCD calculations and its uncertainty determined by the spread of the scaling varying the prediction parameters: quark mass, renormalization and factorization scales.  
The scaling procedure was validated by comparing the results to the existing measurements and to other predictions (MNR, GM-VFNS). 

A clear centrality dependence of the heavy flavor production rate has been observed in PbPb collisions at $2.76$~TeV, as illustrated in Fig.~\ref{fig:ALICEHFPbPb} (right) by the prompt ${\rm D}^0$ mesons ${\rm R}_{\rm AA}$ for $6<p_t<12$~GeV/c. 
The ${\rm R}_{\rm AA}$ transverse momentum dependence of cocktail-subtracted inclusive electrons, inclusive muons, prompt ${\rm D}^0$ mesons, and prompt ${\rm D}^+$ mesons in the most central PbPb interactions
 is presented in Fig~\ref{fig:ALICEHFPbPb} (left). 
The inclusive muon distribution at intermediate $p_t$, mainly populated by heavy flavor decays, shows a suppression of their production rate in good agreement with that of the heavy flavor decay electrons in a different pseudo-rapidity range. 
Prompt charm at mid-rapidity is also suppressed, consistently with heavy flavor leptons and charged pions. 
These results are also in accordance with the CMS measurement of ${\rm J}/\psi$ from beauty decays ${\rm R}_{\rm AA}$~\cite{Dahms:2011QM}. 

The LHC open heavy flavor measurements in PbPb collisions at $\sqrt{s_{NN}}=2.76$~TeV exhibit a decrease of the production rates, relative to the scaled yields in pp collisions, with the reaction centrality, as expected if they lose energy in the created QCD medium. Heavy flavor electrons, muons, and D mesons manifest a similar suppression, compatible also with that of charged pions. Unfortunately, the measurement uncertainties do not allow yet to disentangle if the medium influence differs for light, charm and beauty hadrons, if there is or there is not a mass hierarchy in the energy loss mechanism at LHC energies.  

%
\vspace{-17pt}

\begin{theacknowledgments}
Copyright CERN for the benefit of the ALICE Collaboration. 
\end{theacknowledgments}

\vspace{-15pt}

%


\bibliographystyle{aipproc}   
%
%
%

%

\end{document}